\documentclass[pra,twocolumn]{revtex4}

\usepackage[dvips]{graphicx}%
\usepackage{amsmath}
\usepackage{bm}
\begin{document}
%%%%%%%%
%\small
%%%%%%%%
\title{Optimal Gaussian $N$-to-$M$ cloning with linear optics and Gaussian cloning of known-phase coherent states} %/ Ryo Namiki / \today}

\author{Ryo Namiki}% / \today}
\author{Masato Koashi}
\author{Nobuyuki Imoto}
\affiliation{CREST Research Team for Photonic Quantum Information, Division of Materials Physics, 
Department of Materials Engineering Science, 
Graduate school of Engineering Science, Osaka University, 
Toyonaka, Osaka 560-8531, Japan}
%\date{\today}
\begin{abstract}
 We show how to implement the optimal Gaussian $N$-to-$M$ cloning with linear optics and homodyne detection.  
We also show that the Gaussian $N$-to-$M$ cloning of known-phase coherent states can be performed with the fidelity $\sqrt \frac{2 M N }{2M N+M -N}$ by linear optics and homodyne detection, and with $\frac{2  }{\sqrt{1+\frac{1}{N}}+\sqrt {1-\frac{1}{M}}}$ by utilizing quadrature squeezing.  From the classical limit of the cloning (1-to-$\infty$ cloning), a necessary condition of continuous variable quantum key distribution using known-phase coherent states is provided.
\end{abstract}

%\pacs{03.67.Dd, 42.50.Lc} 
%\keywords{continuous variable, quantum key distribution, Gaussian cloning}
\maketitle

\small
%\newpage
%%%%%%%%%%%%%%%%%%%%%   part 1 %%%%%%%%%%%%%%%%%%%%%%%%%%%%%%%%%%%%%%%%%%%%%%%%
\section{Introduction}
The no-cloning theorem clarifies an interesting distinction between classical and quantum information processing considering the possibilities of making copies of quantum states. Under the name of the cloning, many interesting features in manipulating quantum states have been revealed \cite{rmp-clone}. One of the most familiar setting in cloning problems is the so-called \textit{optimal cloning}, which discusses how well one can make approximate copies of states (with various restrictions). 
The performance of the cloning machines (CM) is estimated by the fidelity between the input and output states. Theoretical goal of the optimal cloning is to show the upperbound of the fidelity and implementation of the CM that achieves the bound.

In continuous-variable (CV) quantum information \cite{CV-RMP}, as accessible optical carrier of information and solvable basic tools, cloning of coherent states \cite{CV-clone} have been investigated extensively. The maximum fidelity of the optimal Gaussian $N$-to-$M$ CM (to make $M$ approximate copies from given $N$ copies of original states, $M>N$) is given by $F_{N\to M } \equiv \frac{MN}{MN +M  -N  } $ \cite{optc1}. Alexanian considered the cloning of coherent states with known phases and proposed an implementation of Gaussian $1$-to-$2$ CM based on four-wave mixing whose achievable fidelity is $\frac{4}{5}$ \cite{kp-clone}.

The restriction of cloning implies the impossibility of noiseless amplification. 
 In fact, the optimal Gaussian cloning can be achieved by using a quantum-noise-limit phase insensitive amplifier \cite{amp}.
The implementation of the Gaussian CM was firstly proposed based on the parametric amplification which requires nonlinear interaction between two optical modes \cite{optc1,optc2}. Recently, it is shown that the quantum-limit amplification can be implemented without nonlinear interaction \cite{amp2} and that the optimal Gaussian $1$-to-$2$ cloning is possible with linear optics and homodyne detection \cite{cv-clone-lo}.

In this paper, we propose the implementation of the optimal Gaussian $N$-to-$M$ CM with linear optics and homodyne detection as a generalization of \cite{cv-clone-lo}.  
We also consider the Gaussian cloning of coherent states with known phases and show an implementation of the $N$-to-$M$ CM that has better performance than the previously proposed one \cite{kp-clone}. %

This paper is organized as follows. In Sec. II, we present the optimal Gaussian $N$-to-$M$ CM with linear optics and homodyne detection. In Sec. III, we present the Gaussian cloning of coherent states with known phases. In Sec. IV, we discuss the necessary condition of CV quantum key distribution using coherent states associated with the classical limit of CMs (the $1$-to-$\infty$ CMs).  
We summarize the results in Sec. V. 
%\begin{widetext}
\begin{figure}[bthp]
\includegraphics[width=1\linewidth]{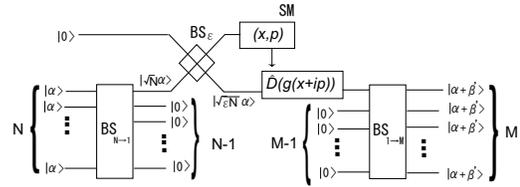}%\includegraphics[width=1\linewidth]{fig2.eps}
\caption{%The optimal $N$-to-$M$ Gaussian cloning machine with linear optics 
\label{fig1} BS$_{\textrm{N}\to \textrm{1},\textrm{1}\to \textrm{M}}$: set of beam splitters, BS$_{\epsilon}$: beam splitter with reflection $\epsilon$, SM: the simultaneous measurement of quadratures, $\beta '$: random noise}%, $\hat D$: the displacement operator
\end{figure}
%\end{widetext}
\section{optimal Gaussian $N$-to-$M$ cloning machine with linear optics}
The scheme of the optimal Gaussian $N$-to-$M$ cloning with linear optics and homodyne detection is shown in FIG. \ref{fig1}.
First, we convert $N$ copies of the input coherent state $|\alpha \rangle$ into $| \sqrt N \alpha \rangle $ using a proper set of beam splitters (BS). Next we split it with a BS that has the reflection $\epsilon$. The output states are $|\sqrt{ \epsilon N}\alpha \rangle $ and $|\sqrt{ (1-\epsilon)N} \alpha \rangle $.
We perform the simultaneous measurement of the quadratures on the state $|\sqrt{(1-\epsilon )N } \alpha \rangle $ by using a double homodyne detector. The measurement is characterized by the $Q$ function of coherent states 
\begin{eqnarray} 
Q_\alpha (\beta)&\equiv& \frac{1}{\pi}\langle \beta | \alpha \rangle\langle \alpha | \beta \rangle = \frac{1}{\pi} e^{-|\beta -\alpha|^2}  \label{Qf}
%Q_\alpha (x,p)&=& \frac{1}{\pi} e^{-(x- X)^2 - (p- P)^2}
\end{eqnarray} where $\beta$ is a complex number whose real part and imaginary part correspond to the normalized outputs of the double homodyne detector \cite{leonhardt}. 
Namely, the probability density that the output of the simultaneous measurement is $\beta$ when the input coherent state is $|\sqrt{(1-\epsilon )N } \alpha \rangle $ is given by 
\begin{eqnarray} 
Q_{\sqrt{(1-\epsilon )N } \alpha} (\beta)
 = \frac{1}{\pi} e^{-|\beta -\sqrt{(1-\epsilon )N } \alpha|^2}.
%Q_\alpha (x,p)&=& \frac{1}{\pi} e^{-(x- X)^2 - (p- P)^2}
\end{eqnarray} 
 Then, we displace the remaining signal $|\sqrt {\epsilon N}\alpha \rangle$ according to the measurement outcome $\beta$ with the gain $g_{N,M} (\epsilon )$. The displaced signal is represented by the density operator \begin{widetext}\begin{eqnarray} 
\hat \rho _{D}&=&     \int Q_{\sqrt{(1-\epsilon )N } \alpha} (\beta) \hat D( g _{N,M} (\epsilon ) \beta ) |\sqrt{ \epsilon N }\alpha   \rangle\langle \sqrt{ \epsilon N }\alpha  | \hat D^\dagger( g _{N,M} (\epsilon ) \beta )  d^2 \beta \nonumber\\
&=& \frac{1}{\pi}  \int  e^{- |\beta |^2 } \big|(\sqrt \epsilon  +\sqrt
{1-\epsilon} g _{N,M} (\epsilon ) )\sqrt N \alpha + g _{N,M} (\epsilon )  \beta  \big\rangle\big\langle (\sqrt \epsilon  +\sqrt
{1-\epsilon} g _{N,M} (\epsilon ) )\sqrt N \alpha + g _{N,M} (\epsilon )  \beta  \big|d^2  \beta ,  %  \sqrt\frac{2}{\pi} e^{-2 x ^2}&=& \frac{1}{\pi}  \int  e^{- |\beta |^2 } |\sqrt M  \alpha + g _{N,M} (\epsilon )  \beta  \rangle\langle \sqrt M  \alpha + g _{N,M} (\epsilon )  \beta |d^2  \beta %  \sqrt\frac{2}{\pi} e^{-2 x ^2}
\end{eqnarray}\end{widetext}
where $\hat D(\beta )$ is the displacement operator which displaces the coherent-state amplitude by the amount $\beta $.
Finally we split $\hat \rho _{D}$ into $M$ parts of equal amplitudes using BSs: \begin{eqnarray} 
\hat \rho &=& \frac{1}{\pi} \int e^{-  |\beta |^2 }  \left| \alpha + \frac{g _{N,M} (\epsilon )  }{\sqrt M }\beta  \Big\rangle \Big\langle \alpha + \frac{g _{N,M} (\epsilon )  }{\sqrt M }\beta \right|^{\otimes M} d^2 \beta , \nonumber \\
%  \sqrt\frac{2}{\pi} e^{-2 x ^2}
\end{eqnarray}
 where we set
\begin{eqnarray} 
g_{N,M} (\epsilon )=\sqrt\frac{M}{(1-\epsilon) N }\left(1- \sqrt\frac{\epsilon N }{M}\right)
\end{eqnarray}
so that the mean amplitude of each clone corresponds to the coherent amplitude of the originals, $\alpha $. Note that the form of $\hat \rho$ is a separable state. 

By tracing out arbitrary $M-1$ modes of $\hat \rho$, we obtain the density operator of each clone: % is given by partial trace of $\hat \rho$ remaining one mode:
\begin{eqnarray} 
\hat \rho _{N \to M}%&\equiv& \textrm {Tr }_{M-1}(   \hat \rho  ) \\
&=&  \frac{1}{\pi} \int e^{-  |\beta |^2 }  \left| \alpha + \frac{g _{N,M} (\epsilon )  }{\sqrt M }\beta  \Big\rangle \Big\langle \alpha + \frac{g _{N,M} (\epsilon )  }{\sqrt M }\beta \right| d^2 \beta . \nonumber \\
%  \sqrt\frac{2}{\pi} e^{-2 x ^2}
\end{eqnarray} 
From this expression, we can see that the quality of the clones becomes better as the additional noise $\frac{g _{N,M} (\epsilon )  }{\sqrt M }\beta $ becomes smaller. The optimal choice of $\epsilon$, which minimizes $g_{N,M}(\epsilon)$, is given by
\begin{eqnarray} 
\epsilon =\frac{N}{M}. \label{ep1}
\end{eqnarray}

Then, using the relation (\ref{Qf}), we obtain the fidelity which achieves the bound:  
\begin{eqnarray} 
   \langle \alpha |\hat \rho _{N \to M} |\alpha \rangle  %\nonumber 
%&=&   \frac{2}{\sqrt{1+4 \Delta x^2 }\sqrt{ 1+ 4 \Delta p^2}}\\  
&=& \frac{MN}{MN +M  -N  }= F_{N\to M } .
\end{eqnarray} 
The quadrature variance of the optimal clones is given by
\begin{eqnarray} 
\Delta x_{N\to M} ^2 =\Delta p_{N\to M} ^2 = \frac{1}{4}+ \frac{1}{2}\left( \frac{1}{N}-\frac{1}{M} \right), \label{qvari}
\end{eqnarray} where we use the normalization of the quadrature variance of coherent states, $\Delta x_0 ^2 \equiv \frac{1}{4}$.

 Note that the measurement operates on the initial $N$ modes coherently. It suggests that the measurement of constructively interfered coherent states is the key element for the optimality when the number of the input states $N$ is more than $1$. On the other hand, the displacement need not operate on the joint modes but can operate on each of the output $M$ modes individually. 
To be concrete, we consider the situation where Alice is given $N$ copies of $|\alpha\rangle$ and her task is to distribute $M$ clones of $|\alpha \rangle$ to remote $M$ parties. In the ``joint'' displacement case, Alice generates the $M$ clones locally and sends each of the clones to each of the remote parties using quantum channels.  
In the individual displacement case, instead of the ``joint'' displacement, Alice sends the coherent states $\{|\frac{N}{M} \alpha \rangle \}$ and the measurement outcome $\beta$ to each of the $M$ parties using quantum channels and classical channels, respectively. Then each of the parties performs the displacement with the gain $g = \frac{g_{N,M}(\frac{N}{M})}{\sqrt M }= \sqrt{\frac{1}{N}-\frac{1}{M}}$ according to the classical information $\beta$. This achieves exactly the same task as the ``joint'' displacement scheme. Although the number of the displacement operations becomes $M$, the amount of each displacement becomes smaller by a factor of $\frac{1}{ \sqrt M}$. %This completes the task and the fidelity is the same as that of the local cloning.

%Note also that the classical limit of the cloning ($M \to \infty$, $\epsilon = 0$) corresponds to a direct measurement on $|\sqrt N \alpha \rangle $. If $N$ is more than $1$, the direct measurement is a joint measurement which coherently operates on the initial $N$ modes.

\section{Gaussian $N$-to-$M$ cloning of known-phase coherent states}

\begin{figure}[htpb]
\includegraphics[width=1\linewidth]{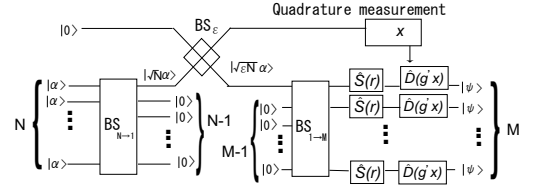}%\includegraphics[width=1\linewidth]{fig2.eps}
\caption{%The optimal $N$-to-$M$ Gaussian cloning machine with linear optics 
\label{fig2}  $\hat S (r)$: Squeezer,  $|\psi \rangle =\hat D (g' x)\hat S (r ) |  \sqrt{\epsilon N /M } \alpha \rangle$: The output state corresponding to the measurement outcome $x$ }%, $\hat D$: the displacement operator
\end{figure}

%\textit{--$N$-to-$M$ Gaussian cloning of known-phase coherent states--}%coherent states with LO}
Let us consider the Gaussian $N$-to-$M$ cloning of known-phase coherent states. Here, the term ``known-phase'' means that the phase-space angle of the coherent-state amplitude is known up to the sign of the amplitude, and so we can set the amplitude $\alpha $ real without loss of generality. 
The cloning procedure is similar to that of the individual displacement case given in the previous section but we utilize the quadrature squeezing and single quadrature measurement (see FIG. \ref{fig2}).%\begin{widetext}
%\end{widetext}

First we transform $N$ copies of the coherent states $| \alpha \rangle$ into $| \sqrt N \alpha \rangle$. % and $N-1$ vacua with a proper set of BSs. 
Next we split  $| \sqrt N \alpha \rangle$ into  $| \sqrt{(1- \epsilon ) N }\alpha \rangle$ and  $M$ parts of  equal amplitudes $|\sqrt{  \frac{\epsilon N}{M}} \alpha  \rangle$ using proper set of BSs. Then, we measure the position quadrature of $| \sqrt{(1- \epsilon ) N }\alpha \rangle$.
In this known-phase case, the clones need not be symmetric in the phase space, so we apply a quadrature squeezing $\hat S (r) $ on each of the remaining $ M$ signals. The squeezing operator transforms the quadratures as \begin{equation}   \hat S^\dagger (r) ( \hat x + i \hat p )  \hat S (r) = \hat x e^{ r}+ i \hat p e^{-r} \label{shat} \end{equation} where $\hat x$ is  the position quadrature and $\hat p$ is the momentum quadrature.
%$| \sqrt{\epsilon N } \alpha \rangle $ which sould be optimized later \hat S (r)   \hat a  \hat S^\dagger (r). 
Finally, we feed forward the measurement outcome $x$ onto each one of the $M$ squeezed states with the gain $g' (\epsilon ) $. This completes the cloning task. 

 The probability density that we obtain the quadrature measurement outcome $x$ is given by the quadrature distribution of the coherent state
\begin{eqnarray} 
|\langle x | \sqrt{(1-\epsilon) N }  \alpha \rangle|^2= \sqrt\frac{2}{\pi}  e^{-2 (x-  \sqrt{(1-\epsilon) N } \alpha )^2}   \label{qd}
\end{eqnarray} where $\langle x |$ is the eigen bra of $\hat x$ with the eigen value $x$.
Then we can write the density operator of each known-phase clone as \begin{widetext}\begin{eqnarray} 
\hat \rho _{N\to M}^{(KP)} &=& \sqrt\frac{2}{\pi}  \int  e^{-2 (x-  \sqrt{(1-\epsilon) N } \alpha )^2} \hat D (g' (\epsilon )  x  ) \hat S (r)  \left|\sqrt\frac{\epsilon N }{M}\alpha     \Bigg\rangle \Bigg\langle \sqrt\frac{\epsilon N }{M} \alpha   \right|\hat S^\dagger (r)  \hat D^\dagger ( g' (\epsilon )  x  )  dx . %\nonumber \\%  \sqrt\frac{2}{\pi} e^{-2 x ^2}
\end{eqnarray} 

The mean value of the position quadrature is given by
\begin{eqnarray}
\langle \hat x \rangle &\equiv& \textrm{Tr}(\hat \rho _{N\to M}^{(KP)} \hat x) \nonumber\\
&=&\sqrt\frac{2}{\pi}  \int  e^{-2 (x-  \sqrt{(1-\epsilon) N } \alpha )^2}\Big \langle \sqrt\frac{\epsilon N }{M} \alpha   \left|\hat S^\dagger (r)  \hat D^\dagger ( g' (\epsilon )  x  ) \hat x   \hat D (g' (\epsilon ) x  ) \hat S (r)  \right|\sqrt\frac{\epsilon N }{M} \alpha     \Big\rangle  dx \nonumber\\
&=&\sqrt\frac{2}{\pi}  \int  e^{-2 (x-  \sqrt{(1-\epsilon) N } \alpha )^2} \Big \langle \sqrt\frac{\epsilon N }{M} \alpha   \left|\hat S^\dagger (r)  ( \hat x + g' (\epsilon )  x  ) \hat S (r)  \right|\sqrt\frac{\epsilon N }{M} \alpha     \Big \rangle  dx \nonumber\\
&=&\Big \langle \sqrt\frac{\epsilon N }{M} \alpha   \Big |  \hat x e^r    \Big |\sqrt\frac{\epsilon N }{M} \alpha     \Big \rangle + g' (\epsilon )   \sqrt\frac{2}{\pi}  \int x e^{-2 (x-  \sqrt{(1-\epsilon) N } \alpha )^2}     dx \nonumber\\
&=& \left( \sqrt\frac{\epsilon N }{M}   e^r  + g' (\epsilon )     \sqrt{(1-\epsilon) N }\right) \alpha ,
\end{eqnarray} \end{widetext} where we use the transformation of the position quadrature by the displacement operator $\hat D^\dagger (x) \hat x \hat D (x) = \hat x +x  $ in the second line and Eq. (\ref{shat}) in the third line. Assuming that the mean quadrature value of the clones corresponds to that of the originals, i.e., $\langle \hat x \rangle  =\alpha $, the gain is determined to be
\begin{eqnarray} 
g' (\epsilon) &=&  \frac{1}{\sqrt{(1-\epsilon) N }}\left(1- e^r \sqrt\frac{\epsilon N }{M}\right).
%=\sqrt{\frac{1}{N}-\frac{e^{2r} }{M }}.
\end{eqnarray}   
The variances of the quadratures are similarly calculated as
\begin{eqnarray} 
\langle(\Delta x)^2 \rangle   &\equiv& \textrm{Tr}(\hat \rho _{N\to M}^{(KP)} \hat x^2) - \textrm{Tr}(\hat \rho _{N\to M}^{(KP)} \hat x) ^2 \nonumber \\ 
&=&\frac{1}{4} \left[ e^{2r}  + g'  ( \epsilon)^2 \right] \\
\langle(\Delta p)^2 \rangle  &\equiv& \textrm{Tr}(\hat \rho _{N\to M}^{(KP)} \hat p^2)  - \textrm{Tr}(\hat \rho _{N\to M}^{(KP)} \hat p) ^2 
= \frac{1}{4}  e^{-2r}.  \label{dpkp}
\end{eqnarray}   
In order to make the noise as small as possible, we minimize $g' (\epsilon)$ by setting 
\begin{eqnarray} 
\epsilon &=& \frac{Ne^{2r}}{M}. \label{ep2}
\end{eqnarray}  
Then, the variance of the position quadrature is represented by a function of $N$, $M$, and $r$: 
\begin{eqnarray} 
\langle(\Delta x)^2 \rangle  &=& \frac{1}{4} \left[ \frac{1}{N} +\left( 1-\frac{1}{M} \right)  e^{2r}  \right] . \label{dxkp}%
\end{eqnarray}

Now we determine the maximum fidelity and the corresponding choice of the squeezing parameter $r$.
By using the relation
\begin{eqnarray} 
\langle\alpha | \hat S (r) | \beta \rangle = \frac{1}{ \sqrt{\cosh r} } e^{% \left( 
-\frac{\alpha ^2 + \beta ^2  }{2 } + \alpha \beta \frac{1}{ {\cosh r} }  + \frac{\alpha ^2 - \beta ^2 }{2} \tanh r  } %\nonumber\\  %\right)   
\end{eqnarray} for real $\alpha$ and $\beta$,  
the fidelity is calculated as \cite{com1} 
\begin{eqnarray} 
F_{N,M}^{(KP)} (r) &\equiv& \langle \alpha |\hat \rho _{N\to M}^{(KP)}|\alpha \rangle \nonumber \\
%&=&    \frac{2}{\sqrt{1+4 \langle(\Delta x)^2 \rangle     } \sqrt{ 1+4 \langle(\Delta p)^2 \rangle   }} \nonumber \\
  &=& \frac{2}{\sqrt{\left( 2+ \frac{1}{N} - \frac{1}{M}\right) +\left( 1+ \frac{1}{N}\right) e^{-2r}+\left( 1- \frac{1}{M}\right) e^{2r} } }\nonumber .\\
 %  &\le& \frac{2}{\sqrt{1+ \frac{1}{N}} +\sqrt{ 1- \frac{1}{M}}} \equiv F_{N \to M}^{(KP)} 
\end{eqnarray} By replacing the last two terms in the square root of this expression with the geometric mean, we obtain %the arithmetric mean of 
\begin{eqnarray} 
F_{N,M}^{(KP)} (r)    &\le& \frac{2}{\sqrt{1+ \frac{1}{N}} +\sqrt{ 1- \frac{1}{M}}} \equiv F_{N \to M}^{(KP)}. 
\end{eqnarray}
The optimal squeezing parameter which achieves the maximum $ F_{N \to M}^{(KP)} $ is given by
\begin{eqnarray} 
e^{2r_\textrm{o} } = \sqrt \frac{(N+1)M}{(M-1)N} .
\end{eqnarray} By inserting $r= r_\textrm{o}$ into Eqs. (\ref{dxkp}) and (\ref{dpkp}), the quadrature variances at the optimal point are given by
\begin{eqnarray} 
\langle(\Delta x)^2 \rangle_{N \to M}^{(KP)}  &=& \frac{1}{4} \left[ \frac{1}{N} +\left( 1-\frac{1}{M} \right)  \sqrt \frac{(N+1)M}{(M-1)N}  \right] %\nonumber\\ 
\\
\langle(\Delta p)^2 \rangle_{N \to M}^{(KP)}  &=& \frac{1}{4} \sqrt \frac{(M-1)N}{(N+1)M}   .
\end{eqnarray}   

The mechanism of the fidelity improvement by the squeezing is simply interpreted as follows.  Although the phase-space asymmetry induced by a squeezing just seems to degrade the fidelity, the squeezing induces another effect. If the squeezing parameter is large, we can adjust the mean quadrature of the clones with relatively small magnitude of the gain. This means that the additional noise imposed by the feed-forward operation becomes smaller, and the smaller noise implies a better fidelity. Therefore, the optimal squeezing parameter is determined by the trade-off between the reduction of the additional noise and loss of the fidelity by the asymmetry.

 Without the squeezing (with linear optics and homodyne detection), the fidelity is given by
\begin{eqnarray} 
F_{N,M }^{(KP)} \left(0 \right)&=& \sqrt\frac{2MN}{2MN +M  -N  }.
\end{eqnarray}
From Eq. (\ref{dxkp}) and (\ref{dpkp}) with $r=0 $, the quadrature variances of the linear optical case become \begin{eqnarray} 
\langle(\Delta x)^2 \rangle  &=& \frac{1}{4} \left(1+\frac{1}{N}-\frac{1}{M} \right),  \\ 
\langle(\Delta p)^2 \rangle   &=& \frac{1}{4}.
\end{eqnarray}

In both cases of the optimal $r=r_\textrm{o}$ and without squeezing $r=0$, the quadrature noise of the clones is not phase insensitive. If one needs the clones that have symmetric noise, from the condition $\langle(\Delta x)^2 \rangle = \langle(\Delta p)^2 \rangle $ in Eqs. (\ref{dpkp}) and (\ref{dxkp}), the squeezing parameter should be 
\begin{eqnarray} 
e^{-2r_* } =  \frac{1+\sqrt{1+4N^2 \left(1-\frac{1}{M}\right)}}{2N}  \label{sr} .
\end{eqnarray}In this case, the fidelity will be
\begin{eqnarray} 
F_{N,M }^{(KP)} \left(r_* \right)&=& \frac{4N}{2N+1 +\sqrt{1+4N^2 \left(1- \frac{1}{M}\right)}  }. %\nonumber \\
\end{eqnarray}
We can verify
\begin{eqnarray} 
F_{N\to M }^{(KP)} 
 \ge F_{N,M }^{(KP)} \left(0 \right)
 \ge F_{N,M }^{(KP)} \left(r_* \right).
\end{eqnarray}
As a comparison with \cite{kp-clone}, we can see that $F_{1, 2}^{(KP)}(r_*) =\frac{4}{ 3 +\sqrt 3} $ is grater than the previous result of the $1$-to-$2$ case, $\frac{4}{5}$. Even the classical-limit fidelity without squeezing $F_{1, \infty}^{(KP)} (0) =\sqrt \frac{ 2}{3} $ exceeds $\frac{4}{5}$. 
The upperbound of the fidelity is unknown and optimality of the present scheme is an open question. %except for the classical-limit (see Sec. IV B). 

%\section{classical limit of cloning}
\section{Necessary condition of quantum key distribution}
The classical limit of the cloning, i.e., measure-and-prepare scheme, makes an entanglement breaking (EB) channel \cite{Bae06}. In the quantum key distribution (QKD), if the observed data is considered to be given from the signal which comes through an EB channel, secret key cannot be distilled. This is because the presence of the EB channel implies that an eavesdropper (Eve) can perform an intercept-resend attack. 

In the experiments of CV QKD, symmetric quadrature noises are observed \cite{hirano03}. In such case, the EB channels that induce symmetric noise provide necessary conditions of CV QKD. If Eve uses the classical-limit cloning of coherent states, the excess noise \cite{namiki2} observed by the legitimate receiver of the QKD becomes \begin{eqnarray}
\delta   \equiv \frac{\langle ( \Delta x^2)\rangle_{\textrm{obs}} - \Delta x_0^2 }{  \Delta x_0^2 } = \frac{   \Delta x _{1\to \infty }^2   - \Delta x_0 ^2}{ \Delta x_0 ^2 } =   2,\nonumber\\ 
\end{eqnarray}where $\langle ( \Delta x^2)\rangle_{\textrm{obs}}$ is the receiver's quadrature variance and we used Eq. (\ref{qvari}). To ensure that Eve does not execute this strategy, the excess noise has to satisfy $\delta < 2 $. Moreover, by putting the clone into a lossy channel with the line transmission $\eta$, Eve makes an EB channel which has the line transmission $\eta $ and excess noise $\delta = 2 \eta $, since the excess noise decreases in proportion to the line transmission. Therefore, the necessary condition of CV QKD using coherent states \cite{namiki2,namiki3} is given by %assuming practical lossy and noisy channel, the excess noise has to satisfy 
\begin{eqnarray}
\delta   <  2\eta .   \label{feq}
\end{eqnarray}

 If we consider CV QKD protocols using known-phase coherent states as in \cite{Rig06,Heid06}, a more stringent necessary condition is provided by the classical limit of the Gaussian known-phase CM.
In the binary phase modulation case \cite{Rig06,Heid06}, the necessary condition is provided by the analysis of the separable condition between a qubit and a mode \cite{Rig06}. The following result can be applied not only for the binary phase modulation case but for any protocol using known-phase coherent states.

From the phase-insensitive-noise case of Eq. (\ref{sr}) with $ N=1$ and $ M\to \infty $, the quadrature variances of Eqs. (\ref{dpkp}) and  (\ref{dxkp}) become 
\begin{eqnarray} 
\langle(\Delta x)^2 \rangle  &=& \langle(\Delta p)^2 \rangle =\frac{1}{4}\frac{\sqrt 5 +1 }{2} 
\end{eqnarray}
Thus, the excess noise by the classical-limit known-phase CM with symmetric noise is determined to be
\begin{eqnarray}
\delta = \frac{\sqrt 5 -1}{2} .
\end{eqnarray}    
Taking into account the line transmission $\eta$ similar to the above \textit{unknown-phase} case, we obtain the relevant necessary condition of CV QKD using known-phase coherent states:
\begin{eqnarray}
\delta < \frac{\sqrt 5 -1}{2} \eta. \label{ffeq}
\end{eqnarray}
This condition is stringent compared with the unknown-phase case of Eq. (\ref{feq}) by more than a factor of $3$. 
%How  It may be interesting to  simply %explains the gap between Eq. (\ref{feq}) and the result of \cite{Rig06}, i.e., in the known-phase case, the eavesdropper need monitor only single quadrature and the excess noise can be suppressed. It suggests that the conditions given by \cite{Rig06} can be relaxed for coherent-state CV QKD using more than binary phase modulations.

\section{summary}
%In conclusion, 
We provided the optimal Gaussian $N$-to-$M$ cloning scheme with linear optics and homodyne detection. We also considered the cloning of known-phase coherent states and provided the implementation of Gaussian $N$-to-$M$ cloning machine. A better fidelity is obtained when the quadrature squeezing is applied. From the classical limit of the known-phase cloning machine that has phase-insensitive noise we found a necessary condition of continuous variable quantum key distribution using known-phase coherent states. %: \begin{equation}\delta < \frac{\sqrt 5 -1}{2} \eta . \nonumber\end{equation} %We have shown that the present Gaussian known-phase cloner with squeezing is the optimal in the classical limit. %In the other cases, 
The bounds of the fidelity and optimality of the present known-phase cloning machines are left for open questions.


\begin{thebibliography}{}  \small
%\bibitem{rmp-qkd}N. Gisin, G. Ribordy, W. Tittel, and H. Zbinden, \rmp \textbf{74,} 145 (2002).


\bibitem{rmp-clone} V. Scarani, S. Iblisdir, N. Gisin, and A. Ac\'in,
\rmp \textbf{77,} 1225 (2005).


%\bibitem{coherent}F. Grosshans and P. Grangier, \prl\textbf{88,} 057902 (2002).
%\bibitem{postsel} Ch. Silberhorn, T. C. Ralph, N. L\"utkenhaus, and G. Leuchs, \prl\textbf{89,} 167901 (2002).
%\bibitem{coherentR}F. Grosshans, G.V. Assche, J. Wenger, R. Brouri, N.J. Cerf, and P. Grangier, Nature 421, 238 (2003).%``Quantum key distribution using gaussian-modulated coherent states'' (London) , 
%\bibitem{no-switch}C. Weedbrook, A. M. Lance, W.P. Bowen, T. Symul, T. C. Ralph, and P. K. Lam, \prl\textbf{93,} 170504 (2004).




%\bibitem{gene}Z. Zhai et al., \pra \textbf{73,} 052302 (2006).

\bibitem{CV-RMP}S. L. Braunstein, and P. van Loock, \rmp  \textbf{77,} 513 (2005).%Quantum information with continuous variables

\bibitem{CV-clone}N. Cerf, A. Ipe, and X. Rottenberg, \prl  \textbf{85,} 1754 (2000).
\bibitem{optc1}S. L. Braunstein, N. J. Cerf, S. Iblisdir, P. van Loock, and S. Massar, \prl \textbf{86,} 4938-4941 (2001).

\bibitem{kp-clone}M. Alexanian, \pra \textbf{73,} 045801 (2006).


%\bibitem{cv-teleclo} P. van Loock and S. L. Braunstein, \prl \textbf{87,} 247901 (2001).
%\bibitem{NMcv-teleclo}J. Zhang, C. Xie, and K. Peng,  \pra \textbf{73,} 042315 (2006). %:  J. Zhang, C. Xie, and K. Peng, \prl \textbf{95,} 170501 (2005)@. 

%\bibitem{noisy-telec} A. Ferraro and M.G.A. Paris, \pra \textbf{72,} 032312 (2005).

%\bibitem{ex-cv-telc} S. Koike et al., \prl \textbf{96,} 060504 (2006).

%\bibitem{conju}  N. J. Cerf and S. Iblisdir, \pra \textbf{62,} 040301(R)  (2000);  N. J. Cerf and S. Iblisdir, \pra \textbf{64,} 032307 (2001). %Phase conjugation of continuous quantum variables

%\bibitem{pci} N. J. Cerf and S. Iblisdir, \prl \textbf{87,} 247903 (2001). %phase congugate input cloning machine 



\bibitem{amp}C. M. Caves, \prd\textbf{26,} 1817 (1982).



\bibitem{optc2} J. Fiur\'asek, \prl \textbf{86,} 4942 (2001).%

\bibitem{amp2}V. Josse, M. Sabuncu, N.J. Cerf, G. Leuchs, and U.L. Andersen, \prl \textbf{96,} 163602 (2006)

\bibitem{cv-clone-lo}U.L. Andersen, V. Josse, and G. Leuchs,  \prl \textbf{ 94,} 240503 (2005).

\bibitem{leonhardt} U. Leonhardt, {\textit{Measuring the Quantum State of Light} } (Cambridge, 1997).
%\bibitem{reverse}R. Filip, J. Fiur\'asek, and P. Marek, \pra \textbf{69,} 012314 (2004).%Reversibility of continuous-variable quantum cloning
%\bibitem{Gro03}F. Grosshans and P. Grangier, Quant. inf. Comp., \textbf{3} 535 (2003); \eprint{ quant-ph/0306141}.


%\bibitem{comment} In Ref \cite{coherentR,Gro03}, the excess noise is defined by $  \epsilon =  \delta /\eta  $.

\bibitem{com1} The fidelity is also simply calculated by  $ 
F \equiv \langle \alpha |\hat \rho |\alpha \rangle \nonumber 
=    \frac{2}{\sqrt{1+4 \langle(\Delta x)^2 \rangle     } \sqrt{ 1+4 \langle(\Delta p)^2 \rangle   }},$ 
which holds for any Gaussian state $\hat \rho $ whose mean amplitude is $\langle\hat  x \rangle +i \langle\hat p\rangle =\alpha$. 
% where $\langle(\Delta x)^2 \rangle $ and $\langle(\Delta p)^2 \rangle $ are the quadrature variances. 
See, e.g., H. Scutaru, J. Phys. A: Math. Gen \textbf{31} 3659-3663 (1998).
  
%\bibitem{namiki4}R. Namiki, M. Koashi, and N. Imoto, \pra\textbf{73,} 032302 (2006).

\bibitem{Bae06} J. Bae and A. Ac\'in, \prl \textbf{97,} 030402 (2006).

\bibitem{hirano03}T. Hirano, H. Yamanaka, M. Ashikaga, T. Konishi, and R. Namiki, \pra \textbf{68,} 042331 (2003).  %; R. Namiki and T. Hirano, \pra\textbf{67,} 022308 (2003).


\bibitem{namiki2}R. Namiki and T. Hirano, \prl\textbf{92,} 117901 (2004).
\bibitem{namiki3}R. Namiki and T. Hirano, \pra\textbf{72,} 024301 (2005).%quant-ph/0506193.


\bibitem{Rig06} J. Rigas, O. G\"uhne and N. L\"utkenhaus, \pra \textbf{73,} 012341 (2006).

\bibitem{Heid06} M. Heid and N. L\"utkenhaus, \pra \textbf{73,} 052316 (2006). 


%\bibitem{Ham05} K. Hammerer, M.N. Wolf, E.S. Polzik, and J.I. Cirac, \prl \textbf{94,} 150503 (2005).

%\bibitem{b92} C. H. Bennett, \prl  \textbf{68,} 3121 (1992).
%\bibitem{dpsk} E. Waks, H. Takesue, and Y. Yamamoto, \pra  \textbf{73,} 012344 (2006).


\end{thebibliography}
\end{document}